\documentclass[preprint,12pt]{elsarticle}

\usepackage{amssymb}
\usepackage{hyperref} 
\usepackage{xurl}
\usepackage{listings}
\usepackage{amsmath}
\usepackage{rotating}
\usepackage{multirow}
\usepackage{xcolor}

\usepackage{subcaption}
\usepackage{graphicx}
\usepackage{todonotes}

\usepackage{tcolorbox} 

\usepackage[normalem]{ulem} 

\journal{Heliyon}

\begin{document}

\begin{frontmatter}



\title{Tripl\'{e}toile: Extraction of Knowledge from Microblogging Text}


\author[1]{Vanni Zavarella\corref{cor1}}
\author[3]{Sergio Consoli}
\author[1]{Diego Reforgiato Recupero}
\author[1]{Gianni Fenu}
\author[1]{Simone Angioni}
\author[2]{Davide Buscaldi}
\author[4]{Danilo Dess\'i}
\author[5,6]{Francesco Osborne}
\cortext[cor1]{Corresponding author}

\affiliation[1]{organization={Department of Mathematics and Computer Science, University of Cagliari},
 addressline={Via Ospedale 72}, 
 city={Cagliari},
 postcode={09121}, 
 country={Italy}}

\affiliation[2]{organization={Laboratoire d'Informatique de Paris Nord, Sorbonne Paris Nord University},
 addressline={99 Av. Jean Baptiste Clement, 93430 Villetaneuse}, 
 city={Paris},
 country={France}}

\affiliation[3]{organization={European Commission, Joint Research Centre (DG JRC)},
 addressline={Via E. Fermi 2749}, 
 city={Ispra (VA)},
 postcode={21027}, 
 country={Italy}}
 
\affiliation[4]{%
 organization={Knowledge Technologies for Social Sciences Department, GESIS Leibniz Institute for the Social Sciences },
 addressline={Unter Sachsenhausen 6-8}, city={Cologne}, postcode={50667}, country={Germany}}

\affiliation[5]{%
 organization={Knowledge Media Institute, The Open University},
 addressline={Walton Hall, Berrill Building}, city={Milton Keynes}, postcode={50667}, country={UK}}

\affiliation[6]{%
 organization={Department of Business and Law, University of Milano Bicocca},
 addressline={Via Bicocca degli Arcimboldi 8}, city={Milano}, postcode={20100}, country={Italy}}

\begin{abstract}

Numerous methods and pipelines have recently emerged for the automatic extraction of knowledge graphs from documents such as scientific publications and patents. However, adapting these methods to incorporate alternative text sources like micro-blogging posts and news has proven challenging as they struggle to model open-domain entities and relations, typically found in these sources.
In this paper, we propose an enhanced information extraction pipeline tailored to the extraction of a knowledge graph comprising open-domain entities from micro-blogging posts on social media platforms. Our pipeline leverages dependency parsing and classifies entity relations in an unsupervised manner through hierarchical clustering over word embeddings. We provide a use case on extracting semantic triples from a corpus of 100 thousand tweets about digital transformation and publicly release the generated knowledge graph. On the same dataset, we conduct two experimental evaluations, showing that the system produces triples with precision over 95\% and outperforms similar pipelines of around 5\% in terms of precision, while generating a comparatively higher number of triples.

\end{abstract}

\begin{graphicalabstract}
\end{graphicalabstract}

\begin{highlights}
\item We design a scalable architecture for triple extraction from social media posts

\item We introduce a relation extraction and mapping method using validated dependency tree patterns and hierarchical clustering

\item We release a knowledge graph comprising 22,270 statements automatically generated from around hundred thousand tweets about digital transformation

\item We show that our pipeline outperforms alternative methods in terms of precision, with a 5\% improvement with respect to similar architectures

\end{highlights}


\end{frontmatter}


\section{Introduction}
\label{sec:intro}
Examining, connecting, and understanding content sourced from microblogging platforms holds significant importance in pinpointing trends, and grasping the intricacies of events and individuals' influence. However, this endeavor is particularly demanding due to the Internet's diverse array of social platforms, each marked by its own distinctiveness, and potentially featuring natural language text in varying formats, structures, and lengths.

Social analysts and various stakeholders commonly navigate this intricate realm through the utilization of social media platforms such as Hootsuite\footnote{\url{https://www.hootsuite.com/}}, Brandwatch\footnote{\url{https://www.brandwatch.com/}}, Talkwalker\footnote{\url{https://www.talkwalker.com/}}, Sprout Social\footnote{\url{https://sproutsocial.com/}}. However, these platforms are constrained to basic queries and merely provide a list of pertinent documents that 
require manual analysis. These limitations represent a notable impediment to the flow of knowledge within the social media analysis process.

The 
main problem lies in the fact that existing systems do not possess an adequate depiction of the nuanced social media dynamics, thereby rendering them incapable of facilitating advanced queries regarding the entities mentioned in the posts. This limitation hinders the ability to discern potential trends, gauge the influence of events or individuals, and understand their relationships.

Consequently, the research community has put forth numerous proposals aimed at generating organized, interconnected, and machine-readable data frameworks of social analysis knowledge found within text from microblogging platforms~\cite{7764416,DORPINGHAUS2022100337,10.1145/3366424.3383112}. Typically, this resulting representation employs Semantic Web technologies, such as ontologies and knowledge graphs. In computer science, ontologies are defined as \textit{explicit specifications of a conceptualization}~\cite{Xiao20185511} 
and serve to formalize the conceptual structure of a specific domain by delineating the categories of entities and their interrelationships. Typically, ontologies are encoded using the Web Ontology Language (OWL) and are considered the foundational pillars of the Semantic Web~\cite{cornerstone2023}. 

Knowledge graphs (KGs) are extensive networks comprising entities and relationships, imparting machine-readable and comprehensible information pertaining to a specific domain, adhering to a formal semantic structure~\cite{DBLP:conf/i-semantics/EhrlingerW16}. In recent years, KGs have become increasingly recognized for their ability to organize structured data in a semantically significant way, allowing them to effectively support various AI systems~\citep{peng2023knowledge}.
The relationship between two entities is typically formalized as a triple 
in the format of \texttt{<subject, predicate, object>}, such as \texttt{<digital transformation, revolutionize, industry>}. The structure of a KG is commonly outlined in a domain ontology. Large-scale KGs are usually generated through a semi-automated process, utilizing both structured and unstructured data. Some prominent examples include DBpedia~\cite{dbpedia2023}\footnote{\url{https://www.dbpedia.org/}}, Google Knowledge Graph\footnote{\url{https://developers.google.com/knowledge-graph}}, BabelNet\footnote{\url{https://babelnet.org/}}, and YAGO\footnote{\url{https://yago-knowledge.org/}}.
Furthermore, knowledge graphs can undergo automatic refinement through link prediction techniques, which are designed to identify additional relationships among domain entities~\citep{kumar2020link,nayyeri2021trans4e}. For instance, these approaches can facilitate the formulation of novel scientific hypotheses by linking known entities in new ways~\citep{borrego2022completing}.

\color{black}Creating extensive and high-quality knowledge graphs from social media is a current open problem that has already been addressed by researchers~\cite{DORPINGHAUS2022100337}. \color{black}Existing solutions either depend on systems that aid social media experts in structuring their knowledge or rely on information extraction pipelines~\cite{webmedia_estendido,DORPINGHAUS2022100337,10.3233/SW-160240,MARTINEZRODRIGUEZ2018339}. The first category of solutions suffer from scalability problems. Information extraction techniques have the potential for scalability but often struggle to generate outputs of sufficient quality for practical applications. Specifically, present approaches for extracting entities and relationships from social media texts typically focus on specific domains~\cite{DORPINGHAUS2022100337} without giving much importance to the preprocessing and linking operations of entities and relations, and their grounding. However, crafting a large-scale, coherent, and semantically sound representation of social media texts drawn from millions of posts is an entirely distinct challenge. Consequently, merely employing existing methods for entity and relationship extraction on an extensive collection of texts would yield a highly noisy outcome~\cite{DESSI2021253}. Therefore, several challenges should be addressed, including:
\begin{itemize}
\item Integrating the extracted information from various posts into a cohesive representation;
\item Evaluating the validity of the resultant triples;
\item Defining a flexible ontological framework to formalize a range of statements originating from social media text.
\end{itemize}

\color{black}
Recent advancements in natural language processing have given rise to sophisticated Large Language Models (LLMs), including Mistral~\cite{jiang2023mistral}, LLaMa 3~\cite{touvron2023llama,huang2024good}, Gemma~\cite{team2024gemma}, and GPT 4.0~\cite{openai2023gpt4}, among others. These models exhibit the ability to generate coherent and articulate responses to user queries and perform various tasks such as text classification and information extraction. Despite their capabilities, concerns have emerged regarding the accuracy and reliability of the content they generate. A notable issue is their tendency to produce ``hallucinations'', i.e., responses that lack grounding in factual knowledge~\cite{xu2024hallucination}. To address this, researchers are exploring the integration of LLMs with structured knowledge representations~\cite{10.1115/1.4052293,siddharth2024retrieval}. This integration aims to enhance the accuracy and transparency of LLMs by linking them to reliable sources and enabling the tracking of claim origins. KGs are increasingly vital in this context and are well-suited to complement LLMs~\cite{peng2023knowledge}.\color{black}

\color{black}Similar challenges for KG construction have \color{black}already been addressed within the scholarly domain in~\cite{DESSI2021253} where the authors introduced an information extraction approach merging data from various tools based on a domain ontology and allowing in this way for the creation of expansive KGs. This pioneering approach has served as a source of inspiration for subsequent research in the field~\cite{cit1,cit2,cit3,cit4,cit5}. However, this 
work also encountered several limitations: i) the entity extraction modules did not capitalize on the expert knowledge gained from analyzing the resulting knowledge graphs; ii) limited capability to unify multiple instances of the same entity; iii) a shallow and manual approach for mapping verbal predicates to semantic relations; iv) a constrained methodology for evaluating triple validity, relying on a basic multilayer perceptron classifier.

Therefore, in this paper, we present \textit{Tripl\'{e}toile}, an enhanced information extraction architecture designed to overcome the aforementioned limitations. This innovative solution demonstrates the capability to extract entities from social media text and identify different instances of them. Additionally, it facilitates the extraction of various relationships among these entities by using hierarchical clustering, word embeddings, and dimensionality reduction techniques. Furthermore, we present a use case consisting of the application of the proposed architecture to a subset of around 100k tweets extracted from the Twitter platform\footnote{\url{https://twitter.com/}} from 2022 and concerning the digital transformation domain.

We conducted an assessment of Tripl\'{e}toile by comparing it to several alternative solutions using a benchmark dataset consisting of 500 triples. As it will be shown next, our results reveal that Tripl\'{e}toile outperforms the alternatives in terms of accuracy, while at the same time generating a relatively higher number of triples. 

In brief, the main research contributions of this paper encompass the following:
\color{black}
\begin{itemize}
 \item We design a general, scalable, and flexible architecture for triple extraction from social media text.
 \item We provide a use case on Twitter where we extracted approximately 100k tweets related to digital transformation in 2022 and subsequently released a corresponding knowledge graph comprising 22,270 statements.
 \item On the proposed use case, we perform a formal assessment of Tripl\'{e}toile in terms of precision and a comparative evaluation with respect to alternative methods.
 \item We publicly release the resulting triple store as a dataset 
 within the Joint Research Centre Data Catalogue\footnote{\url{https://data.jrc.ec.europa.eu/dataset/f7be47f7-49a2-44e8-9dc8-043735af4139}}, as well as within the European Data portal\footnote{\url{https://data.europa.eu/88u/dataset/f7be47f7-49a2-44e8-9dc8-043735af4139}}, the official data repository of the European Commission.
\end{itemize}
\color{black}

The remainder of this paper is organized as follows. The related work is illustrated in Section~\ref{sec:relatedwork}. The proposed architecture is depicted in Section~\ref{sec:methodology}. The use case and comprehensive analysis of the extracted triples are described in Section~\ref{sec:usecase}. The evaluation we have carried out, including the comparisons against state-of-the-art tools, is detailed in Section~\ref{sec:evaluation}. Finally, conclusions and future works where we are headed are reported in Section~\ref{sec:conclusions}.

\section{Related Work}
\label{sec:relatedwork}

The term ``knowledge graph'' was first coined in 1972, but it was not until 2012 that it gained widespread recognition after Google's announcement\footnote{\url{https://blog.google/products/search/introducing-knowledge-graph-things-not/}} of the Google Knowledge Graph~\citep{Hitzler202176}. This event also sparked the growth of knowledge graph development and usage in the industry~\citep{Ji2022494,Noy2019}. A knowledge graph is a graph of data that is designed to capture and communicate knowledge about the real world. Its nodes represent entities of interest and its edges represent the relationships between these entities~\citep{peng2023knowledge,Hogan2021}.

Creating, maintaining, and refining knowledge graphs requires the use of an array of techniques for information extraction, entity selection and linking, relation extraction, and ontology engineering~\citep{Ristoski20161,cornerstone2023,Tudorache2020125}. Numerous scholarly articles delve into the methodologies for generating knowledge graphs across different domains and under various constraints.~\citep{DESSI2021253,Dessi2020127}. 
Notably, \citet{Sequeda2019526} introduced a unique pay-as-you-go approach to overcome the challenges associated with understanding complex database schemas, providing a use case from a large company.

The extraction of knowledge graphs from web sources to answer questions related to social networks~\cite{7764416}, such as Twitter or Facebook, has been widely discussed in literature~\citep{Collarana2018359, Gabrilovich20161195,DORPINGHAUS2022100337}. 
~\citet{10.1145/3366424.3383112} described how to build knowledge graphs for social networks by developing deep Natural Language Processing models, and holistic optimization of knowledge graphs and the social network. 
While authors in~\cite{Choudhary20211373} have already acknowledged the overlap between social networks and knowledge graphs, the current research has poorly exploited this overlap so far. 
A number of information extraction pipelines have been proposed to create high-quality knowledge graphs within the social network analysis domain (see for example~\citep{webmedia_estendido,DORPINGHAUS2022100337,10.3233/SW-160240,MARTINEZRODRIGUEZ2018339}). While information extraction techniques have the potential for scalability, they often struggle to produce outputs of sufficient quality for practical applications. Specifically, current approaches for extracting entities and relationships from social analysis texts typically focus on specific domains~\cite{DORPINGHAUS2022100337}, neglecting the significance of preprocessing, linking operations, entity grounding, and the creation of a large-scale, coherent, and semantically robust representation of social network analysis texts drawn from millions of posts~\cite{DESSI2021253}.

\citet{Haslhofer2018} have emphasized the importance of connected knowledge graphs and discovery, whereas~\citet{Hyvonen2019230} have highlighted the significance of new relationships extracted from the original dataset. 
In recent years there has been also an increasing research focus on ontologies and interoperable data~\cite{Cristofaro2021}.
In particular,~\citet{DESSI2021253} have proposed an information extraction method that combines data from different tools using a domain ontology, enabling the creation of expansive knowledge graphs. This first approach has been a source of inspiration for further research in the field~\cite{cit1,cit2,cit3,cit4,cit5}.

Implicitly, a significant amount of research has already utilized knowledge graphs. This involves combining actors, persons, and additional information such as locations using linked data~\citep{Mountantonakis2019}. While the practice of using external data linked to a network is still prevalent, it is also possible to define different types of nodes. \citet{cit4} conducted a systematic review of the process of knowledge graph creation. The review methodology aimed to collect and describe the various steps involved in this process, such as data identification, construction of the knowledge graph ontology, knowledge extraction, analysis of the extracted knowledge, knowledge graph creation, and maintenance. The last step, maintenance, entails periodic updates and edits to keep the knowledge graph up to date. 
In their review, the authors offer suggestions, best practices, and tools that support the creation and maintenance of knowledge graphs.

In this paper, we propose a methodology specifically tailored for micro-blogging text which overcomes several limitations of existing approaches in the field.
Specifically: i) our method identifies entities among the text and has a high ability to unify different instances 
of the same entity; ii) the designed entity extraction modules make use of the information acquired from analyzing the obtained knowledge graph; iii) we perform entity coreference resolution by applying pronoun anaphora resolution and a set of heuristics to normalize the identified entities; iv) the method recognizes relationships among the identified entities and comes up with an automated approach for mapping verbal predicates to semantic relations; v) finally, a robust methodology to evaluate the validity of the produced triples is adopted.
To the best of our knowledge, a methodology specifically tailored for micro-blogging text embracing all these features is the first of its kind.

\section{The Proposed Architecture}
\label{sec:methodology}
\begin{figure}
\centering
\includegraphics[width=1.1\textwidth]{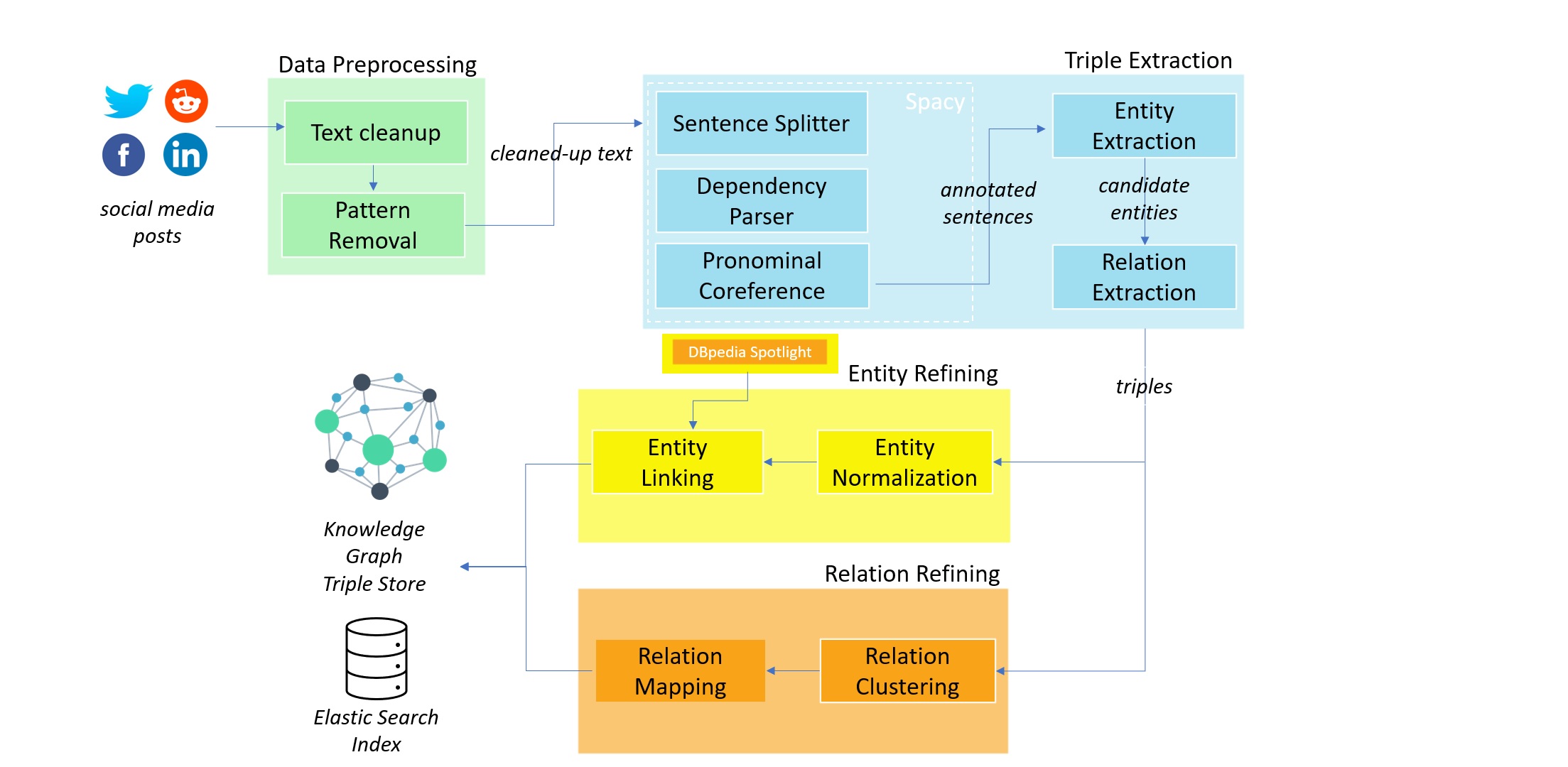} 
\caption{Flowchart of the pipeline for generating a knowledge graph from micro-blogging text data.}
\label{fig:architecture}
\end{figure}

Figure~\ref{fig:architecture} shows the workflow of the pipeline that we propose in this paper.

The main blocks of the architecture include:
\begin{itemize}
 \item \emph{Data Preprocessing}, a step responsible for the normalization of the micro-blogging text in order to make it processable by the downstream text analysis modules;
 \item \emph{Triple Extraction}, the block comprising core modules applying text processing libraries and models for the extraction of entity-relation triples;
 
 \item \emph{Entity Refining}, a block responsible for the cleaning and generalization of entity mentions to canonical forms, in view of subsequent entity merging;
 \item \emph{Relation Clustering}, in which relation instance verbal forms are mapped to canonical forms, computed as a representative of the relation cluster they belong to.
\end{itemize}
The final output of the pipeline is a knowledge graph of generalized triples annotated with references to the micro-blogging text items they were matched in.

The following subsections describe in more detail the individual components of the pipeline across the four main blocks and how they are applied. 

\subsection{Data Preprocessing}
\begin{figure}
\centering
\begin{subfigure}{1.0\textwidth}
\centering
\includegraphics[width=1.0\textwidth]{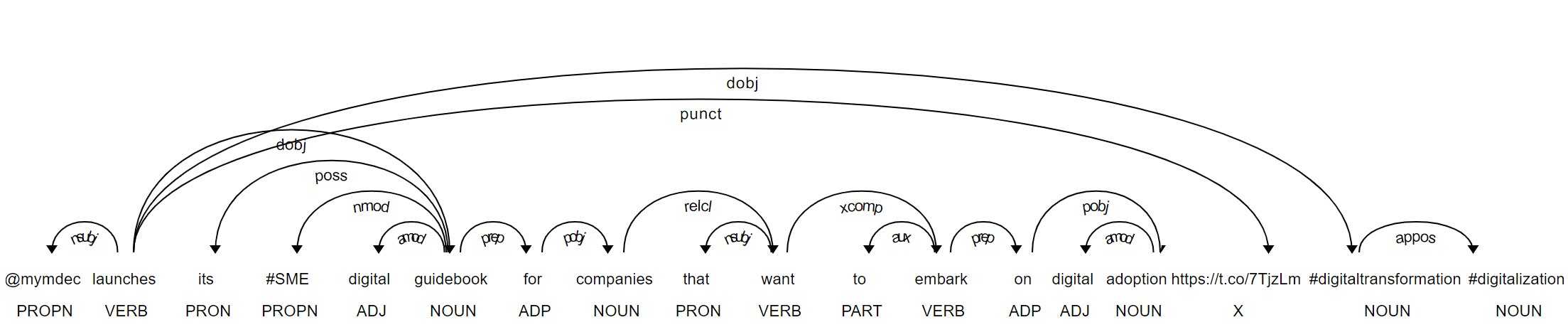}
 \caption{Dependency parse of a tweet's original text.}
 \label{fig:first}
\end{subfigure}
\hfill
\begin{subfigure}{1.0\textwidth}
\centering
\includegraphics[width=1.0\textwidth]{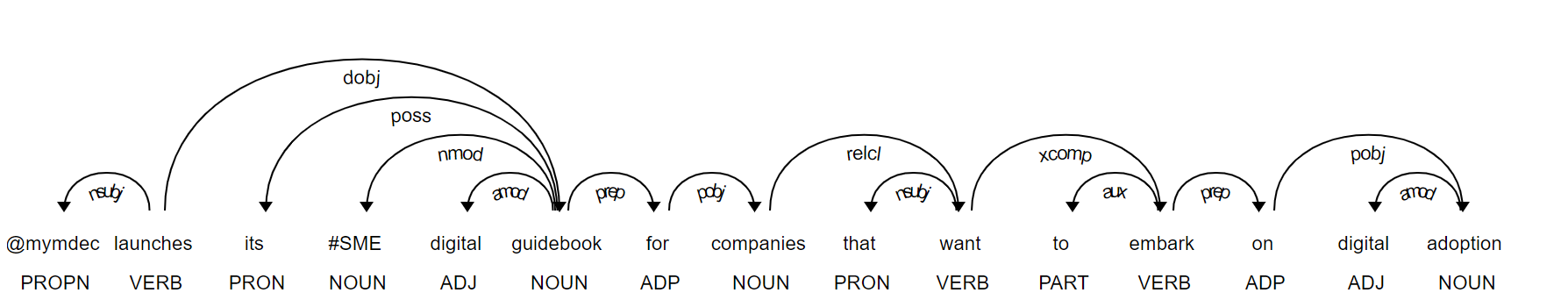}
 \caption{Dependency parse of a tweet after preprocessing}
 \label{fig:second}
\end{subfigure}
\hfill 
\caption{Example of tweet preprocessing.}
\label{fig:tweet preprocessing}
\end{figure}

Twitter status updates (tweets) are short micro-blogging posts of a maximum of 280 characters: their informal (often plainly ungrammatical) genre and the abundance of platform-specific conventions are known to be hard to process by standard NLP tools.
Prior to extracting triples, \color{black}we follow a two-fold approach to tweet normalization~\cite{Siddharth_Blessing_Luo_2022} which can be readily extended to normalize social content from other platforms~\cite{10.1115/1.4029562,CHIARELLO2020103299}. \color{black}
On the one hand, we remove tokens and token sequences encoding platform-specific metadata or denoting communicative conventions that (typically) do not carry any syntactic function in the tweet sentence. Namely, we remove: 
\begin{itemize}
 \item sentiment emoticons and smileys;
 \item reserved tokens (e.g., RT for `retweet');
 \item URLs.
\end{itemize}
On the other hand, we keep by default other platform-specific tokens that can carry syntactic functions depending on the context like hashtags and @ entity mentions (e.g. \textit{\#digitaltransformation, @NASA}). Then, we identify token patterns that typically disrupt the syntactic parsing of the sentence, and remove them from the original tweet. Namely, we implemented the following preprocessing heuristics:
\begin{enumerate}
 \item we remove sequences of $n$ entity mentions and retweet markers at the beginning of a sentence, with $n > 1$ or when the sequence is not followed by a verb. For example, we remove the leading sequence in \textit{``@bansijpatel @RTatsat @kiranpatel1977 Thanks for updating the information with us.''} but not in \textit{``@AMDRyzen enabling \#DataAnalytics in [...]''}.
 \item for any sequence of size $n > 1$ hashtags/mentions/URL, we drop the sub-sequence with indexes $[1:n]$ or drop the entire sequence if preceded by a sentence closing marker like ('!',':','?','.'). For example, in the text \textit{``According to the @PayNews survey, 84 percent of \#employees in the U.S. have instant access to \#information about their pay and \#benefits \#Sapper \#AI \#hr \#support \#goals[...]''} we keep only the first element of the trailing hash tag sequence. 
 \item we remove a leading sequence of $n$ non-verbal tokens ($n$ empirically set to 6) ended by a column sign (e.g. \textit{`Tech Update: Apple Watch's data `black box’ poses research problems [...]'}) as these frequent constructs (carrying a function similar to a tweet title) tend to mislead the dependency parsing.
\end{enumerate}
Entity mentions and hashtags, that are typically removed from tweet preprocessing pipelines for NLP tasks such as sentiment analysis, are highly relevant for knowledge graph generation as they can be nominal subjects, objects, or modifiers of dependency parse trees and therefore be extracted as elements of candidate triples, like the tokens @mymdec and \#SME in Figure~\ref{fig:first}. Notice, although, that the trailing sequence of purely referential elements can often lead to noisy edges, for example in the figure the parser wrongly draws a $dobj$ dependency edge from the main verb ``launches'' onto the hashtag \#digitaltransformation.

Figure~\ref{fig:second} shows that the application of the preprocessing heuristics above (rule 2 in this case) can enhance the parsing on the tweet, without losing too much information\footnote{Although we are currently not using them for KG generation, we are currently saving each tweet's metadata.}.
The preprocessing step is carried out using the output of Spacy's English transformer pipeline \textit{en\_core\_web\_trf-3.6.1} after customizing the default Tokenizer in order to parse tweet metadata (e.g., mentions and hashtags)\footnote{\url{https://github.com/explosion/spacy-models/releases/tag/en\_core\_web\_trf-3.6.1}}.

\subsection{Triple Extraction}
In the triple extraction block, preprocessed tweets are split into sentences and each sentence is fed to the Spacy pipeline mentioned above.
Building upon the works in \cite{DBLP:conf/semweb/DessiORBM22} and \cite{DBLP:journals/kbs/DessiORBM22}, we define a set of procedures to extract candidate nominal entities and predicative triples connecting them from Spacy dependency parse trees.

\subsubsection{Entities}
The entity extraction module detects local nominal phrases with a restricted range of syntactic modifications (e.g., compound nouns, and adjectives). Then it connects and expands them with:
\begin{itemize}
 \item a non-recursive set of attached prepositional phrases;
 \item Spacy quantity-type entities ({\sc MONEY}, {\sc PERCENT}, {\sc QUANTITY}, {\sc CARDINAL}).
\end{itemize}
We also use pronominal anaphora links output by the Spacy pipeline component coreferee\footnote{\url{https://github.com/richardpaulhudson/coreferee}} and assign to it the expanded entity span of the token it points to.

Overall, the module ends up with a set $E = {e_0,...,e_n}$ of non-unified, candidate entity phrases.

\begin{figure}
\centering
\includegraphics[width=1.0\textwidth]{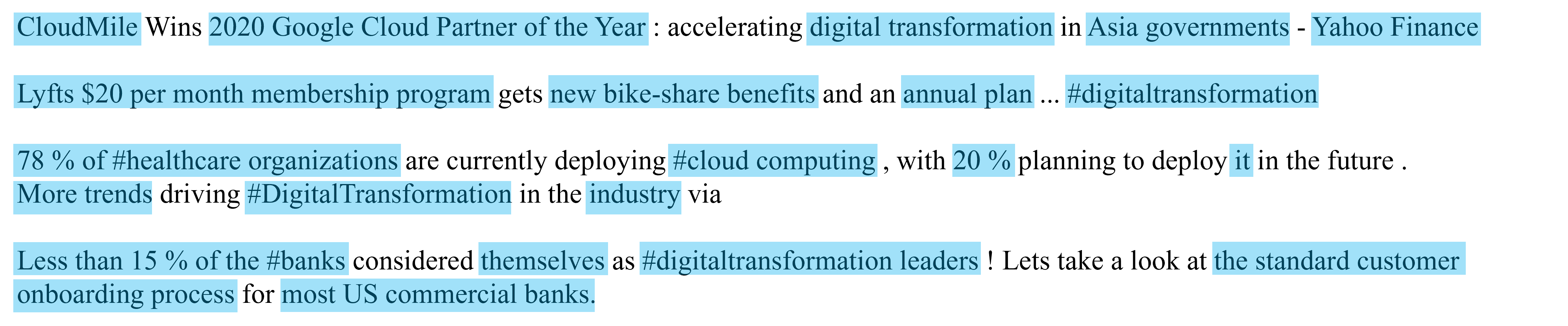} 
\caption{Visualization of candidate entities extracted from a sample of tweets.}
\label{fig:SampleEntities}
\end{figure}

In Figure~\ref{fig:SampleEntities} we show a sample of extracted candidate entities.
For multi-token entity spans including quantifying modifiers (e.g. \textit{`Less than 15\% of the \#banks'}) we maintain a structured representation separating the lexical head (\textit{`\#banks'}) from the quantifying modification of the noun phrase (\textit{`Less than 15\%'}), which then allows a more accurate entity normalization (see Section \ref{Entity Refining} below).

Notice that at this stage the hashtag \textit{\#digitaltransformation} in the second sentence and the noun phrase \textit{digital transformation} in the first are not mapped to the same general concept {\sc digital transformation}23 yet, so that the triples in which they occur would still be considered as unrelated.

\subsubsection{Relations}
\label{relations}
In the relation extraction module, for each sentence $s_i$ all the shortest paths of the dependency tree between each pair of entities $(e_{m},e_{n})$ containing a verb and matching any of the patterns listed in Table~\ref{TargetPaths} below are selected. 

The target pattern set has been selected through an expert validation process. 
First, an open-domain text corpus was automatically annotated with entities, and all shortest paths connecting any pair of entities were collected, resulting in a total of approximately 15k path instances, ranging over 3695 path patterns. The patterns were then sorted by their frequency in the corpus. 
Next, the top twenty patterns, with frequencies ranging from 79 to 1098, were manually reviewed. Three independent evaluators assessed a random sample of 20 triples from each pattern.
Specifically, each evaluator was tasked to assess the correctness of all 400 relevant triples.

In order to be annotated as valid, a triple should reflect the semantics of the portion of the sentence where it was extracted. For example, the triple $<Mr.\ Lewis; give; quixotic\ guided\ tour>$ extracted from the sentence \textit{`Mr. Lewis gives the reader a quixotic guided tour through Silicon Valley while showing how its success stories revolutionized American business.'} with path $[nsubj, dobj]$ was considered valid by the annotators. 
On the other hand, the triple $<air;	rising; hot\ day>$ from the sentence \textit{`Howe says it was discovered by cows drawn to cool air rising from the mouth of the cave on a hot day .'}, with path $[acl, pobj]$ \textcolor{black}{was discarded as most of the annotators did not believe it accurately reflected the semantics of the corresponding text.}

A majority vote was used in order to label each triple as correct/incorrect and only the subset of patterns with a prevalence of correct triples (i.e., more than 10) were considered reliable and kept in the result list.

Although this pattern expert validation process was carried out on a separate text corpus, while evaluating our pipeline on micro-blogging posts (see Section~\ref{sec:evaluation}), we noticed that a potential source of noise was the extraction of triples via the dependency path $[acl,dobj]$. 
The issue arose in instances where the noun's clausal modifier was an infinitive verb, as exemplified in the following sentence:

\textit{`Salesforce really has the power to transform your business.'}

from where a triple $<power,transform,business>$ was wrongly extracted. Consequently, we added a constraint to the dependency path $[acl,dobj]$ in order to filter out those paths where verb nodes had a relation \textit{aux} with an infinitive particle node. In the example above, \textit{transform} has an \textit{aux} relation to the particle \textit{to} and, therefore, it is discarded. More in detail, the following expressions hold:

\textbf{\url{SUBJ}}=\textit{power} $\xrightarrow{\text{acl}}$ \textbf{\url{PRED}}=\textit{transform} $\xrightarrow{\text{dobj}}$ \textbf{\url{OBJ}}=\textit{your business}

\textbf{\url{PRED}}=\textit{transform} $\xrightarrow{\text{aux}}$ \textbf{\url{to}}.

The entire updated process generates a set of verbal relations $V = {v_0,...,v_k}$ and a set of triples $S = {s_0,...,s_k}$ of the form $<e_{m},v,e_{n}>$ where $v\in V$ and $e\in E$.

\begin{table}
\centering
 \begin{center}
 \begin{small}
 \begin{tabular}{|lc|} 
 \hline
 \multicolumn{2}{|c|}{Target dependency paths}\\\hline
 $[nsubj,dobj]$ & \\
$[acl,relcl,dobj]$ &\\
$[acl,dobj]$ &\\
$[nsubjpass,agent,pobj]$ &\\
$[nsubj,dobj,conj]$ &\\
$[nsubj,conj]$ &\\\hline
 \multicolumn{2}{|c|}{Sample discarded paths}\\\hline
 $[obj,pobj]$ &\\
 $[obl,pobj]$ &\\
 $[nsubj;pobj;nmod]$ &\\ 
 \hline
 \end{tabular}
 \end{small}
 \end{center}
 \vspace{-0.4cm}
 \caption{List of target and some of the discarded relation dependency paths.} 
 \label{TargetPaths}
\end{table}

Analogously to what we pointed out for the entities, note that $v$ in $V$ are surface forms, that is individual inflected verbal forms that are unable to generalize triples over morphological or lexical variations. So for example the following triples:

$<BLEND360,acquires,Engagement Factory>$

$<BLEND360,acquired,Engagement Factory>$

$<BLEND360,bought,Engagement Factory>$ 

are considered distinct at this stage.

The final goal of the pipeline is to allow to generalize from the set $S = {s_0,...,s_k}$ of surface form triples of type $<e_{m},v,e_{n}>$, to the lower sized set $T = {t_0,...,t_h}$ of triples of the form $<\epsilon_{m},r,\epsilon_{n}>$ where each $\epsilon_{i} \in E$ is an entity and $r$ is a label in a common relation vocabulary $R$.


\subsection{Entity Refining}
\label{Entity Refining}
The function of this block is to clean up and normalize the candidate entities into a normalized form that allows the merging across entity name variants\footnote{Splitting is another typical subtask of Entity Refining functions, for example by separating the individual entities in parsed coordinated noun phrases like in \textit{`\#testautomation and \#datamanagement can accelerate your \#digitaltransformation'} We deal with these cases earlier on at the triple extraction phase by generating a triple for each coordinated entity.}.

Entities are first cleaned up by removing leading/trailing punctuation marks as well as stop-words.
Afterwords, we distinguish the following cases for normalization:
\begin{itemize}
 \item \textcolor{black}{For hashtags and @ mentions, we remove hashtags and @ symbols, split the ``camel case'' forms (e.g., \textit{\#SmartCities}) and lowercase the resulting string.}
 \item For all other entities, we lemmatize and lowercase all component tokens whose POS tag is neither Verb nor Proper Noun, otherwise we simply lowercase.
 \item For nouns that have variants in American English, we finally map to the British English variant. 
\end{itemize}

We take advantage of such normalized versions of candidate entities in order to merge them, by using the Spacy DBpedia Spotlight Entity Linking library\footnote{\url{https://spacy.io/universe/project/spacy-dbpedia-spotlight}}. 


The DBpedia Spotlight model is trained to perform both entity detection and linking. In order to power this module with the entity normalisation performed by our pipeline, we run it on modified tweet sentences where the original subjects and objects entity spans (extracted by the Entity Extraction module of the Triple Extraction block in Figure~\ref{fig:architecture}) are replaced with their normalized forms. Next, we link the normalized versions of the entities to the corresponding DBpedia entries of the Spacy native entities whose text spans are both:

\begin{itemize}
 \item included within the subject or object text spans of the corresponding normalized version of the entities;
 \item overlapping with the syntactic heads of the corresponding normalized version of the entities. 
\end{itemize}

In other terms, we let the Spacy DBpedia Spotlight module perform the merging of entities that were normalized to the same or similar forms, by having them linked to the same DBpedia unique entries. 
For example, the two candidate entities \textit{`Gartner'} and \textit{`@Gartner\_inc'} are merged 
together by linking them (later formalized with a relation \texttt{owl:sameAs} to the DBpedia entry of the Gartner consulting firm {\url{http://dbpedia.org/resource/Gartner}).

In case only the first condition is met, we assign a semantic `relatedness' link between the candidate entity and the DBpedia entry, indicating that the former is not an instance of, but rather related to the latter\footnote{We keep out the cases when only the second condition is met, as they typically arise from inaccuracies of the entity span detection.}. 
For example, the span \textit{`@gartner\_survey'} is considered only `related' (later mentioned with a relation \texttt{skos:related}) to the DBpedia entry for Gartner.

In Section \ref{sec:usecase} we describe how these relations are encoded in the resulting knowledge graph by inheriting from existing ontology relations.

\subsection{Relation Refining}

This block aims to find the best predicate label $r$ for each relation verb $v$ in a triple $<e_{m},v,e_{n}>$ and to map $v$ to $r$ in the resulting triple.

The approach we followed consisted of deriving a word embedding representation of the verb predicates from a pre-trained model, computing an optimized clustering of the relation vectors, and finally using a representative instance of each cluster to map verb predicates.

After experimenting with several (contextual and non-contextual) word embedding models and clustering algorithms, we converged to a setup using static word embeddings learned with GloVe (Global Vectors for Word Representation,~\cite{pennington2014glove}) and applying HDBSCAN clustering to the vectors. We tested using verb phrase contextual embeddings from Huggingface's \textit{bert-large-uncased}\footnote{https://huggingface.co/bert-large-uncased} and Sentence-BERT\footnote{\url{https://sbert.net/}}. However, it turned out that the optimal cluster scores, in this case, were achieved for a number of clusters too close to the number of items in the dataset\footnote{In other terms, these representations were not suitable for generalizing enough over relations, probably due to the context-specific information they are encoding.}. \color{black}In \ref{appendixA} we report the clustering scores and number of resulting clusters for some best performing configurations using the different embedding models.

\color{black}
\paragraph{Relation Embeddings}
For each single or multi-token relation predicate, we get the static, 300-dimensional word embedding vector made available for text Span objects in the Spacy \textit{en\_core\_web\_lg-3.6.0} pipeline\footnote{\url{https://github.com/explosion/spacy-models/releases/tag/en\_core\_web\_lg-3.6.0}}.

\paragraph{Dimensionality Reduction and Clustering}
We used the HDBSCAN clustering algorithm enhanced by previously applying UMAP dimension reduction technique on the word embeddings vectors\footnote{https://umap-learn.readthedocs.io/en/latest/parameters.html}. 
HDBSCAN is a hierarchical version of the popular density-based DBSCAN algorithm, which is characterized by considering outliers and leaves unclustered the data points lying in low-density regions \cite{Malzer2020223}. Consequently, high dimensional data require more observed samples to produce the suitable level of density for HDBSCAN to work properly. 
However, applying UMAP to perform non-linear, manifold aware dimension reduction \cite{mcinnes2020umap} has been proven to transform the datasets down to a dimension small enough for HDBSCAN to cluster the vast majority of instances.

In order to optimize the combination of UMAP and HDBSCAN, we perform a grid search over the hyper-parameters of both algorithms and evaluate the clustering using the score indicated in Equation~\ref{eq1}:
\textcolor{black}{
\begin{equation}
S = {silhouette_X}\cdot{clustered_X},
\label{eq1}
\end{equation}
}

where the silhouette coefficient $silhouette_x$ of an instance $x \in X$ is defined in Equation~\ref{eq2}:

\begin{equation}
(b – a) / max(a, b),
\label{eq2}
\end{equation}

with $a$ being the mean distance to the other instances in the same cluster and $b$ being the mean distance to the instances of the next closest cluster. 

In the $S$ score formula, the $silhouette_X$ is the mean silhouette coefficient over all the instances of the dataset $X$ that were actually clustered by HDBSCAN \cite{Batool2021} 
while $clustered_X$ is the fraction of instances of $X$ that were actually clustered by HDBSCAN. 

In practice, we optimize for the classical measure of cluster cohesion and separation while penalizing the configurations with low coverage of the dataset.
We finally chose a subset of best-scoring hyper-parameter configurations and plotted their $S$ score over the number of output clusters they generate, so that we are able to pick a sub-optimal configuration that balances between generalization (fewer clusters) and accuracy (cluster number closer to the dataset size).

\paragraph{Relation Mapping}
\textcolor{black}{The triples in our dataset often contain numerous distinct relations that might be seen as synonyms. For instance, relations such as ``includes'', ``involves'', ``embeds'' and ``contains'' can convey similar meanings. To minimize redundancy and support semantic retrieval of the triples in the graph, we consolidate these extracted relations into a smaller set of predefined relations.
Therefore,} for each relation verb $v$ in the dataset, we replace it with the predicate label $r$ consisting of the lemma of the most frequent relation in the cluster of $v$. Otherwise, we map it to itself if $v$ was an outlier and not clustered. Thus, the three distinct triples shown in the last example of Section~\ref{relations} would be merged and the resulting triple would be: 

$<BLEND360,BUY,Engagement Factory>$\footnote{\textcolor{black}{A CSV file with a sample of the most frequent normalized triples, together with the originally matched relations can be found in the project repository at \url{https://github.com/zavavan/dtm_kg/blob/master/data-collection/twitter/sampleNormalizedTriples.tsv}}}.

\textcolor{blue}{}

\section{The Use Case: Digital Transformation Monitoring}
\label{sec:usecase}

The recent surge in data science and artificial intelligence technologies has led to significant insights and aided in the creation of numerous decision-making tools \citep{DS4EF2021}. These instruments assist investors in decision-making and policymakers in creating policy interventions, which have the potential to boost economic growth and enhance societal well-being \citep{taddy2019,marwala}.

In particular, the application of these cutting-edge technologies to social media and news has great potential since they provide a larger set of information than standard lower frequency socio-economic indicators \citep{barbaglia2022,consoli2022}.

These opportunities and challenges are inspiring the research activities at the European Commission's Competence Center on Composite Indicators and Scoreboards\footnote{European Commission's Competence Center on Composite Indicators and Scoreboards (COIN): \url{https://composite-indicators.jrc.ec.europa.eu/}.} at the Joint Research Centre (JRC)\footnote{The Joint Research Centre (JRC) of the European Commission (EC): \url{https://ec.europa.eu/info/departments/joint-research-centre_en}.} aimed at the development of a tracker of societal and economic activities in European countries 
using unconventional data \citep{Colagrossi2022323}.

In light of this, we have deployed our prototype pipeline to develop a Digital Transformation monitoring system from alternative sources.
The technological domain of Digital Transformation is widely pervasive in both scholarly and industrial publications (scientific papers, patents) as well as in the fast-reactive news and social media, capturing the latest updates in the field: therefore, it represents a relevant benchmark for the capacity of our prototype to link and extend existing knowledge graphs generated from conventional sources.
At this aim, we have generated a knowledge graph in the domain of Digital Transformation from a topic-specific tweet dataset.

The dataset was collected by using the Twitter public API v2 full-archive search endpoint, retrieving English language tweets from 2022 containing the hashtag \#DigitalTransformation. We excluded all retweets. 
From the approximately 4M tweets matching the query, we sampled a dataset of around 100k\footnote{After removal of duplicates and near-duplicates, namely tweets over a 0.85 Levenshtein string similarity threshold, computed after preprocessing.} and run the pipeline on it.

The resulting DTSMM\_KG (Digital Transformation Social Media Monitor Knowledge Graph) comprises approximately 22,270 statements. We represented all claims extracted from the tweets using the DTSMM\_KG ontology we created for this purpose\footnote{The ontology definitions are located within the same file of the triple store}. Table \ref{SampleTriples} shows a sample of these statements. 
The reader notices that we refer to statements or triples indifferently. The triples obtained after the reification have not been taken into account for the statistics reported in this paper.

We reified each claim into \textit{dtsmm-ont:Statement} class instances, where \textit{dtsmm-ont} is the namespace prefix of the DTSMM\_KG ontology and a \textit{dtsmm-ont:Statement} defines a specific claim extracted from a given number of tweets. Namely, each statement includes:
\begin{itemize}
\item the reification of the triple itself via \textit{rdf:subject},\textit{rdf:predicate} and \textit{rdf:object} predicates;
\item a data property \textit{dtsmm-ont:hasSupport} reporting the number of tweets supporting the claim;
\item a number of object property instances \textit{dtsmm-ont:comesfromTweet} ranging over ontology instances of type \textit{dtsmm-ont:Tweet} (which was inherited from \textit{schema:SocialMediaPosting}) supporting the claim;
\item A boolean data property \textit{dtsmm-ont:negation} flagging whether a negation of the claim's predicate was parsed from the source text. 
\end{itemize}

Figure~\ref{fig:StatementReification} shows a shortened example of a claim reification having the DTSMM\_KG ontology's instance \textit{machine\_learning} as \textit{rdf:object} and support equal to six.

\begin{figure}
\begin{flushleft}
\begin{verbatim}
dtsmm-ont:statement_10100 a dtsmm-ont:Statement,
 rdf:Statement ;
dtsmm-ont:negation false ;
dtsmm-ont:comesfromTweet dtsmm:tweet_1424266328882429952 ;
...
dtsmm-ont:hasSupport 6 ;
rdf:subject dtsmm:multi_page_document_classification ;
rdf:predicate dtsmm-ont:use ;
rdf:object dtsmm:machine_learning .
\end{verbatim} 
\end{flushleft}
 \caption{A shortened example of reification for a Statement concerning the instance \textit{machine\_learning}, grounded by 6 tweets, with the three dots referring to the hidden \textit{dtsmm-ont:comesfromTweet} predicates.}
 \label{fig:StatementReification}
\end{figure}

In Figure~\ref{fig:ML_Subgraph} instead we visualize a sub-graph of DTSMM\_KG showing a few sample triples having the instance \textit{machine\_learning} as the subject. Here, we report just the statements, hiding claim reification for the sake of readability.

\begin{figure}
 \centering
 \includegraphics[width=0.9\textwidth]{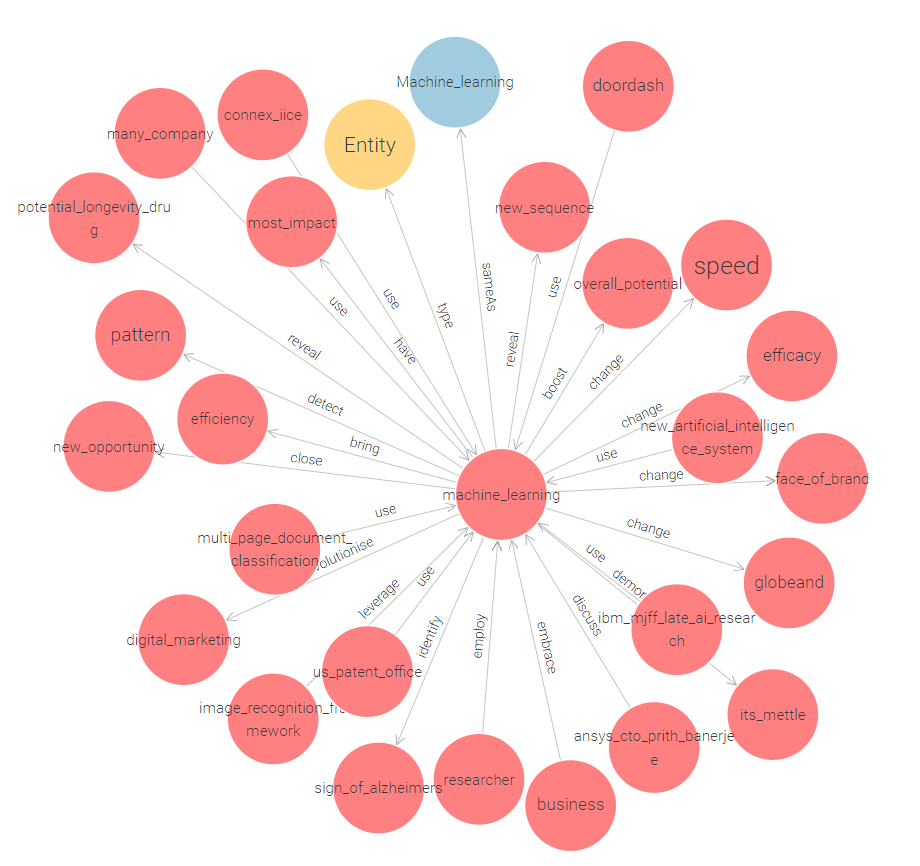}
 \caption{A subgraph from DTSMM\_KG showing a few sample claims for the instance \textit{machine\_learning}, prior to applying explicit statement reification.}
 \label{fig:ML_Subgraph} 
 \hfill
\end{figure}

The linking of the DTSMM\_KG instances to the DBpedia ontology is implemented by using the \textit{owl:sameAs} and \textit{skos:related} predicates in order to encode entity equality and relatedness, respectively. DTSMM\_KG provides 2,857 \textit{owl:sameAs} links
and 3,309 \textit{skos:related} links to to DBPedia entries. 
Figure~\ref{fig:ML_EntityLinkingSubgraph} shows some examples \textit{owl:sameAs} and \textit{skos:related} edges from a number of entities onto the DBpedia resource \url{http://dbpedia.org/resource/Machine\_learning} (the node in yellow). 

The resulting data have been made publicly accessible under Creative Commons Attribution 4.0 International (CC BY 4.0) license\footnote{Creative Commons Attribution 4.0 International license: \url{https://creativecommons.org/licenses/by/4.0/}} within the Joint Research Centre Data Catalogue\footnote{\url{https://data.jrc.ec.europa.eu/dataset/f7be47f7-49a2-44e8-9dc8-043735af4139}}, as well as within the European Data portal\footnote{\url{https://data.europa.eu/88u/dataset/f7be47f7-49a2-44e8-9dc8-043735af4139}}, the official data repository of the European Commission. The direct link to the 
Digital Transformation knowledge graph, available in Terse RDF Triple Language (Turtle), is \url{https://jeodpp.jrc.ec.europa.eu/ftp/jrc-opendata/CC-COIN/se-tracker/DTSMM_KG.ttl}. 
\color{black}In Appendix B 
we also explain how the knowledge graph can be used in domain applications, along with a practical Q$\slash$A exercise via Retrieval Augmented Generation (RAG) \cite{RAGseminal}.
\color{black}

We now report on the experimental results of generating the DTSMM\_KG knowledge graph via the processing modules described in Section \ref{sec:methodology}.

\begin{figure}
 \centering
 \includegraphics[width=0.9\textwidth]{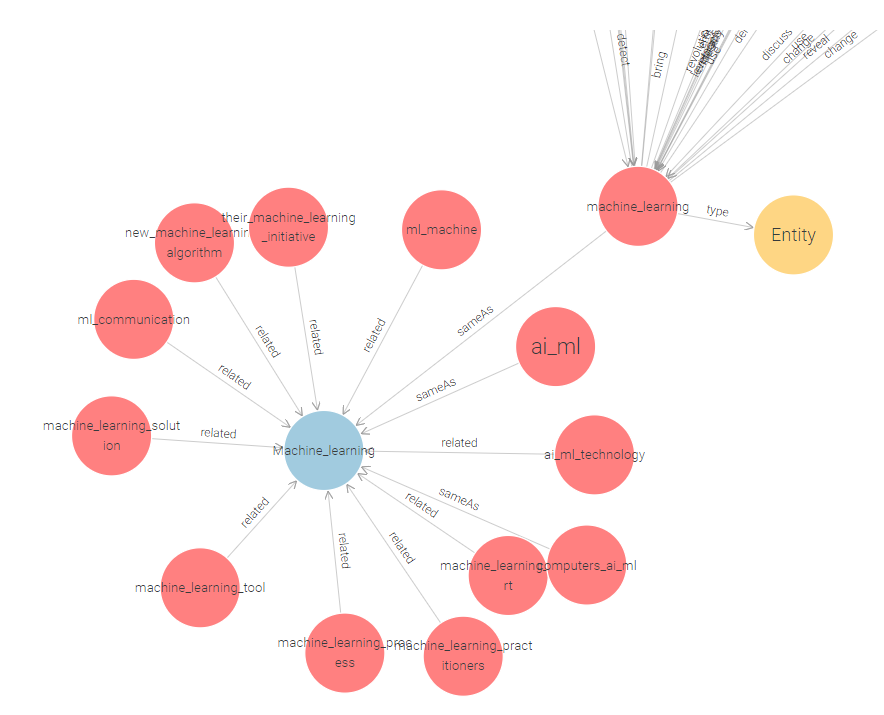}
 \caption{A subgraph from DTSMM\_KG showing entity linking via \textit{owl:sameAs} and \textit{skos:related} predicates of some KG instances to the DBpedia resource \url{http://dbpedia.org/resource/Machine\_learning} (the blue node).}
 \label{fig:ML_EntityLinkingSubgraph} 
 \hfill
\end{figure}

\begin{table}
\centering
 \begin{center}
 \begin{small}
 \begin{tabular}{c|c|c} 
 \hline
 Subject Entity & Relation & Object Entity \\ \hline
 pandemic & accelerate & digital\_transformation \\
 artificial\_intelligence & impact & insurance\_sector \\
 microsoft & buy & riskiq \\
 data-driven\_insight & drive & decision-making \\ 
agile\_business & demand & effective\_marketing\_capability\\
 hootsuite & buy & ai\_chatbot\_firm \\
 automl & generate & data-driven\_insight \\
 image\_classification & use & transfer\_learning \\
 new\_belgium\_brewing & implement & digital workflow\_place\_solution \\
 e-rupi & back & existing indian rupee \\
82\%\_of\_cio & implement & new\_technology \\
image\_recognition\_framework & use & artificial\_intelligence \\
microinsurance & close & africa\_insurance\_gap\\
hsbc\_qatar & introduce & mobile\_payment \\
ford\_motor\_company & explore & blockchain\_technology \\
\hline
 \end{tabular}
 \end{small}
 \end{center}
 \vspace{-0.4cm}
 \caption{A sample of statements extracted by our pipeline.} 
 \label{SampleTriples}
\end{table}

\subsection{Preprocessing and Triple Extraction}



As for the extracted entities, around $33.9\%$ and $6.44\%$ included hashtags and @ entity mentions, respectively; $3.34\%$ were complex noun phrases with prepositional attachments while around $16.6\%$ contained quantitative modifiers of any type (currency, percent, etc.).

Out of all the generated triples, a $5.98\%$ had either the subject or object entity made by a resolved pronominal anaphora, while $0.93\%$ and $4\%$ had $NEGATION$ and $INTERROGATIVE$ tags, respectively.

\subsection{Relation Clustering}
Starting with a set of 29,335 raw triples\footnote{These are surface-level candidate triples from the Triple Extractor, counted prior to entity and relation merging.}, we derived 2,539 unique 300-dimensional word embeddings from GloVe and standardized them.

Via the grid search optimization described in Section \ref{sec:methodology}, we converged to a UMAP two-dimensional representation of the vector dataset, using a $n\_neighbors = 5$ hyper-parameter, which constrains UMAP to look at rather local neighborhoods of five data points when attempting to learn the manifold structure of the data. 
We then optimized UMAP dimensionality reduction and HDBSCAN clustering on this reduced dataset via the hyper-parameter grid search described in Section \ref{sec:methodology}.

Table \ref{RelationMappingSample} shows some sample mappings from relations to their associated predicate labels, consisting of the lemma of the most frequent relation in their clusters.

\begin{table}
\centering
 \begin{center}
 \begin{small}
 \begin{tabular}{|c|c|p{6cm}|} 
 \hline
 Relation Verb & Relation Predicate & Example \\ \hline
 
 fuel & FUEL & \textit{`How the UR+ Ecosystem is Fueling Cobot Market Growth'}\\
 driven by & FUEL & \textit{`Digital transformation in Ho Chi Minh is being driven by remote working'}\\
 accelerated by & FUEL & \textit{`huge social trends being accelerated by the pandemic.'}\\
 identify & IDENTIFY & \textit{`Machine learning can identify signs of Alzheimers in patients '}\\
quantify & IDENTIFY & \textit{`Research quantifies G's potential in roaming and manufacturing '}\\
predict & IDENTIFY & \textit{`AI-supported test can predict eye disease that leads to blindness'}\\\hline
 \end{tabular}
 \end{small}
 \end{center}
 \vspace{-0.4cm}
 \caption{Sample relation verb-predicate mapping.} 
 \label{RelationMappingSample}
\end{table}


\section{Experimental Evaluation}
\label{sec:evaluation}

We perform a twofold evaluation of our methodology. \textcolor{black}{First, we evaluate the precision, recall, and F1 by manually assessing the truthfulness of a test set of statements}. Second, we evaluate our pipeline's precision against a number of alternative tools. 

\subsection{Human Expert Assessment}
For the first evaluation, we randomly selected 483 statements, equally distributed among high-support (support greater or equal to 5) and low-support triple groups. 
Each triple was assessed by three evaluators as True or False. The `True' label was assigned only when all the following criteria were satisfied:
\begin{itemize}
\item{the subj and obj entities are linked by a relation in the tweet text;}
\item{the assigned relation label entails the relation verb in the tweet text;}
\item{the spans of the subject/object of extracted triples include the syntactic head of the relation's subject/object\footnote{For example, a triple $<78\%\_of\_\#healthcare,USE,Digital\_Transformation>$ would be marked as False if extracted from the text \textit{`78\% of \#healthcare organisations deploy \#DigitalTransformation'} as the syntactic head is organisations.}}.
\end{itemize}

We calculated the average pair-wise Cohen $\kappa$ inter-rater agreement, resulting in a value of 0.61. This value generally suggests a significant level of agreement. 
We also computed the Fleiss $\kappa_F$ agreement of all the 3 raters. This is ranging in $[-1,+1]$ and is defined according to~\cite{Falotico2015463} as: 
\begin{equation}
 \kappa_F = \dfrac{p_o - p_e}{1-p_e},
\end{equation}
where $p_o$ is the observed inter-annotator agreement and $p_e$ is the prior probability estimates of the inter-annotator agreement, that is the agreement that we would expect if the annotators were annotating randomly.
The Fleiss $\kappa_F$ score reaches 0.558, which indicates a substantial agreement, although one annotator featured an outlier rating on a specific category of cases.


We evaluated the 483 statements extracted by our pipeline using the majority vote of the three annotators, yielding a precision of 0.96. \textcolor{black}{To compute the recall, the three annotators were assigned an additional task: extracting triples that they deemed correct from the same tweets containing the 483 selected statements. The total count of these tweets was 491, which exceeded the number of triples due to some being extracted from multiple tweets. The total number of triples manually extracted from the annotators was 484 (we considered the union of all the triples extracted by each annotator). Consequently, we were able to calculate the number of true positives (TP), false positives (FP), and true negatives (TN). Table~\ref{PrecisionRecallFMeasure} displays the TP, FP, and TN values for the 484 manually extracted triples. The table shows that the achieved recall was 0.95 and the F1 score was 0.95.}

\begin{table}[hbt!]
\centering
 \begin{center}
 \begin{small}
 \begin{tabular}{|c|c|c|c|c|c|} 
 \hline
 TP & FP & TN & Precision & Recall & F1\\ \hline
 464 & 19 & 20 & 0.96 & 0.95 & 0.95 \\
\hline
 \end{tabular}
 \end{small}
 \end{center}
 \vspace{-0.4cm}
 \caption{Triple evaluation over a manually annotated set of 491 tweets.} 
 \label{PrecisionRecallFMeasure}
\end{table}

Individual rater estimates ranged from 0.90 to 0.96. \color{black}Overall, these results indicate that the pipeline can extract triples with good precision from noisy text like tweets, while at the same time missing only a few triples.

\color{black}
Upon analyzing the results, we identified the primary error sources in the following descending order: i) 
failure in the syntactic parsing of the sentence (5 cases), ii) inaccuracy of relation clustering/mapping (4 cases), and iii) error in pronominal anaphora resolution (4 cases).

\subsection{Comparative Evaluation}
In the second evaluation, we randomly sampled 500 tweets from the 100k-sized original dataset and used our pipeline to extract candidate entities. We then merge this set of entities with the one generated by the DyGIEpp Extractor \cite{wadden-etal-2019-entity}.

DyGIEpp is an IE framework that is able to jointly extract a set of 6 pre-defined types of entities (Method, Task, Material, Metric, Other-Scientific-Term, and Generic). To detect the entities DyGIEpp uses a feed-forward neural network on textual span representations and computes a score for each entity type; an entity is detected considering the highest score for an entity type if a minimum threshold is met.

We then employed four alternative methods to identify relationships between these entities and thus extract the statements from the 500 tweets. Specifically, we compared:

\begin{itemize}

 \item \textbf{OpenIE Extractor}, the IE tool of the Stanford Core NLP suite~\cite{angeli-etal-2015-leveraging}, which is used to extract open-domain relationships composed by only one verb among candidate entities from our pipeline;

 \item \textbf{PoST Extractor}, a module built on top of the Stanford Core NLP suite that uses PoS tags to find all verbs that exist between two candidate entities in a sentence to extract verb relations, using a window of max token distance 15 between the entities;

 \item \textbf{Dependency-based Extractor}, a module that extracts dependency trees using the dependency parser of the Stanford Core NLP suite, maps entities previously extracted using DyGIEpp into the sentence tokens, and exploits $12$ hand-crafted paths\footnote{\url{https://github.com/danilo-dessi/SKG-pipeline/blob/main/resources/path.txt}} to find verbs that connect entities.

 \item \textbf{Entity and Relationship Refiner}, a module that applies \textit{Entity and Relationship Handlers} as described in~\cite{DBLP:journals/kbs/DessiORBM22} to the set that includes \textit{OpenIE Extractor}, \textit{PoST Extractor}, and \textit{Dependency-based Extractor} triples. Its resulting triples have normalized entities that underwent several preprocessing steps, and the relationships are mapped to a controlled vocabulary which ensures that extracted verbs with a similar meaning are mapped to the same relationship.
 
\end{itemize}

The number of extracted triples from the dataset ranged from 339 for the Dependency-based Extractor to a maximum of 1,015 for the PoST method. The latter is a quite predictable outcome as the PoST Extractor combines type-restricted and open-domain entities and at the same time extracts as candidate relations all the verbs occurring between any pair of entities in text, without filtering on the dependency relations connecting them.

After PoST Extractor, our Tripl\'{e}toile pipeline is the one generating the largest number of triples (663) among the methods using the extended range of candidate entities, with OpenIE Extractor producing 588 triples Entity and Relationship Refiner reaching approximately 348 triples.
\textcolor{black}{In order to use these numbers as an indirect assessment of the relative recall levels of the different pipelines, we manually assessed also the precision on a limited random sample of 150 triples generated by each method\footnote{Notice that these test sets are not generated from the same tweet subset for each pipeline. Notice also that the random sampling was done without using any information on the triple support, which was not available for the alternative pipelines.}.}


\textcolor{black}{In order to evaluate the precision of these tools against Tripl\'{e}toile, we manually assessed such 150 triples produced by every technique\footnote{\textcolor{black}{Note that this test set is not generated from the same tweet subset, for each pipeline; moreover, the random sampling was done without using any information on the triple support, which was not available for the alternative pipelines.}}.}

Similarly to the previous evaluation, three evaluators reviewed each triple as `True' or `False'. The annotators' agreement reached a $\kappa_F$ of 0.86, indicating a strong agreement. Finally, we calculated the precision of the five methodologies using the majority vote.
We report the results in Table \ref{ComparativeEvaluation}.

While not as high as in the previous test set, the precision of our pipeline on this smaller sample largely outperforms all the alternative methods. 
Interestingly, it also yielded a significant advantage over the Dependency-based Extractor method, which deploys very similar syntactic information from the sentence. This may be due to the application of the processing step upstream of the triple extraction process.

\begin{table}
\centering
 \begin{center}
 \begin{small}
 \begin{tabular}{|c|c|} 
 \hline
 Extraction Method & Precision \\ \hline
 OpenIE Extractor & 0.52\\
 PoST Extractor & 0.17\\
 Dependency-based Extractor & 0.77\\
 Entity and Relationship Refiner & 0.31\\
 Tripl\'{e}toile & \textbf{0.82}\\
\hline
 \end{tabular}
 \end{small}
 \end{center}
 \vspace{-0.4cm}
 \caption{Precision (P) of the triples extracted from a set of alternative methods from a set of 500 tweets, using a combination of Tripl\'{e}toile and DyGIEpp candidate entities.} 
 \label{ComparativeEvaluation}
\end{table}

\section{Conclusions and Future Works}
\label{sec:conclusions}
In this paper, we presented Tripl\'{e}toile, an information extraction architecture optimized to generate open-domain knowledge graphs from micro-blogging text. The method is mostly unsupervised and does not integrate information from a target domain during the extraction process.
Nonetheless, in a topic-specific test collection of tweets related to the domain of Digital Transformation, the pipeline proved to outperform some of the state-of-the-art methods, generating mostly valid triples. Moreover, we showed that around 12\% of entity occurrences are linked to DBpedia entries, suggesting that the method is potentially useful for tracking relevant entities in the target social media text collection.

\color{black}We are currently experimenting with the transferability of the pipeline across different target domains and preliminary results are promising. As an example, we deployed it for the extraction of a significantly larger graph of 431k triples in the domain of Digital Health and found out that a 8\% of the 86k extracted entities could be linked to DBpedia entries of domain relevant types (e.g., \textit{dbpedia:Disease}, \textit{dbpedia:Company}, \textit{dbpedia:Drug}).
Moreover, the pipeline runtime proved to scale linearly with the size of the document set, which in this case consisted of a larger corpus of 95k news items (23.8M words against 2.9M words of the current tweet collection).

\color{black}A current limitation of our method is that it does not rely on the ontology specification of a target domain in order to customize the entity and relation extraction process. 
As a consequence, extracted entities are currently un-typed, which does not support the execution of more structured queries. Moreover, a domain-specific classification schema for relations would allow setting up a supervised learning of the relation mapping. We expect this to benefit from fine-tuning contextual word embedding representations using Transformer architectures.

Therefore, we plan to work on an enhanced version of the pipeline that builds upon the entity and relation spans generated with the current approach and further classifies them into a granular representation tailored to the specific domain.

As a longer-term goal, this will also help analyzing more thoroughly the usage of social media and other dynamic information sources for tracking and expanding existing knowledge graphs generated for scientific and technological domains.

Last but not least, in light of the recent emergence of numerous scalable large language models, we intend to explore their potential to improve triple extraction methods~\citep{pan2023unifying}. Simultaneously, we aim to capitalize on the resultant knowledge graph to develop knowledge plugins~\cite{meloni2023integrating}, thus augmenting the proficiency of these language models across various natural language processing tasks.

\section*{Acknowledgements}
We acknowledge financial support under the National Recovery and Resilience Plan (NRRP), Mission 4 Component 2 Investment 1.5 - Call for tender No.3277 published on December 30, 2021 by the Italian Ministry of University and Research (MUR) funded by the European Union – NextGenerationEU. Project Code ECS0000038 – Project Title eINS Ecosystem of Innovation for Next Generation Sardinia – CUP F53C22000430001- Grant Assignment Decree No. 1056 adopted on June 23, 2022 by the Italian Ministry of University and Research (MUR).

\bibliographystyle{elsarticle-num-names} 
 \bibliography{biblio}

\newpage

\color{black}\appendix

\section{Relation Clustering Performance for Different Embeddings} \label{appendixA}

\begin{table}[htp!]
\centering
 \begin{center}
 \begin{small}
 \begin{tabular}{|c|c|c|} 
 \hline
 Embedding Model & Silhouette $\cdot$ Clustered Ratio & Num Clusters \\ \hline
 BERT & 0.9387 & 1107 \\
BERT & 0.9287 & 918 \\
 BERT & 0.9171 & 1063 \\
 Sentence-BERT & 0.6852 & 869 \\
 Sentence-BERT & 0.6794 & 978 \\
 Sentence-BERT & 0.6767 & 1050 \\
 GloVe & 0.6505 & 327 \\
 GloVe & 0.6362 & \textbf{332} \\
 GloVe & 0.6345 & 511\\
\hline
 \end{tabular}
 \end{small}
 \end{center}
 \vspace{-0.4cm}
 \caption{\color{black}The table presents clustering score values and the number of output clusters for the top three performing UMAP-HDBSCAN configurations across three tested embedding models. It's worth noting that the dataset comprises a total of 29,335 relation instances for contextualized BERT and Sentence-BERT embeddings. In contrast, for static GloVe embeddings, we consolidated single occurrences of each relation form, resulting in a final set of 2,539 relations due to their context-independent vector representations.} 
 \label{alternativeEmbeddings}
\end{table}

\section{Example of Use in Domain Applications via Q$\slash$A and RAG}}\label{appendixB}

\color{black}
The knowledge graph we have developed is directly applicable to various domain applications, particularly within the realm of Digital Transformation monitoring. Our approach bridges the gap between the wealth of information available in real-time data streams, like social media, and more static, conventional sources. This fusion yields a dynamic and comprehensive view of the Digital Transformation landscape, aiding in real-time monitoring, informed decision-making, and predictive analytics. For instance, the European Commission’s Competence Center on Composite Indicators and Scoreboards at the Joint Research Centre (JRC) is at the forefront of exploring unconventional data to track societal and economic activities across European countries. This activity aligns with our efforts and showcases a practical application where our knowledge graph can significantly contribute. We have utilized our prototype pipeline to create a monitoring system specifically tailored to the domain of Digital Transformation. This technological area is not only prevalent in academic and industrial literature, such as scientific papers and patents but is also actively discussed in dynamic platforms such as news outlets and social media, which often provide the most current insights. The pervasive nature of Digital Transformation makes it an excellent domain for demonstrating the utility of our knowledge graph.

Moreover, the knowledge graph might also serve as a critical resource for enriching Retrieval Augmented Generation (RAG) models \cite{RAGseminal}. In detail, RAG models combine the power of language models with a retrieval component, and our knowledge graph can be used as a novel RAG approach to fetch relevant information during the generation process. By querying our knowledge graph, a RAG model can dynamically pull in contextual data related to Digital Transformation, thus enhancing the quality and relevance of its outputs.

In the following, we provide a Q$\slash$A exercise showing how the knowledge graph can be used in domain applications via RAG. Suppose for example that you wish to know whether the multinational Microsoft is making significant investments in Computer Security. One might supply the following question to a RAG system:\\

\textit{Is Microsoft dedicating substantial resources to computer security technologies?}\\

\newpage
Using a Named-Entity Recognition model\footnote{See e.g. \url{https://huggingface.co/search/full-text?q=named-entity+recognition&type=model}}, the system is able to recognize the entities \textit{Microsoft} and \textit{Computer Security} from the text.

Our knowledge graph, referred to as \textit{DTSMM\_KG}, can be queried via SPARQL to detect whether \textit{Microsoft} entities are declared into its ontology:

\begin{figure}[hb!]
\begin{flushleft}
\begin{tcolorbox}
\begin{footnotesize}
\begin{verbatim}
SELECT DISTINCT *
FROM <DTSMM_KG> 
WHERE { <http://dtsmmkg.org/dtsmmkg/resource/microsoft> ?p ?o . }
\end{verbatim}
\end{footnotesize}
\end{tcolorbox}
\end{flushleft}
 \label{fig:Query1}
\end{figure}

which would produce the following resulting triples (in RDF Turtle format): 

\begin{figure}[hb!]
\begin{flushleft}
\begin{tcolorbox}
\begin{footnotesize}
\begin{verbatim}
@prefix dtsmm: <http://dtsmmkg.org/dtsmmkg/resource/> .
@prefix dtsmm-ont: <http://dtsmmkg.org/dtsmmkg/ontology#> .
@prefix owl: <http://www.w3.org/2002/07/owl#> .

dtsmm:microsoft a dtsmm-ont:Entity ;
 owl:sameAs <http://dbpedia.org/resource/Microsoft>,
 <http://dbpedia.org/resource/Xbox_Live> .
\end{verbatim}
\end{footnotesize}
\end{tcolorbox}
\end{flushleft}
 \label{fig:Query2}
\end{figure}

From this we learn that the \textit{Microsoft} resource is defined and exists into our KG, and also that it is the well-known DBpedia entity \url{http://dbpedia.org/resource/Microsoft}, which would allow us to infer additional information, external to our knowledge-base, via the DBpedia SPARQL endpoint\footnote{Available at \url{https://dbpedia.org/sparql}} with query:

\begin{figure}[hb!]
\begin{flushleft}
\begin{tcolorbox}
\begin{footnotesize}
\begin{verbatim}
SELECT DISTINCT *
FROM <DTSMM_KG> 
WHERE { <http://dbpedia.org/resource/Microsoft> ?p ?o . }
\end{verbatim}
\end{footnotesize}
\end{tcolorbox}
\end{flushleft}
 \label{fig:Query3}
\end{figure}

which would produce 960 knowledge triples about Microsoft (to see all the different triples, you can browse directly DBpedia to \url{http://dbpedia.org/resource/Microsoft}).

This existing knowledge can be enriched with the relations extracted from our Digital Transformation knowledge graph, via the SPARQL query:

\begin{figure}[h!]
\begin{flushleft}
\begin{tcolorbox}
\begin{footnotesize}
\begin{verbatim}
prefix dtsmm: <http://dtsmmkg.org/dtsmmkg/resource/> 
prefix rdf: <http://www.w3.org/1999/02/22-rdf-syntax-ns#> 

SELECT DISTINCT *
FROM <DTSMM_KG> 
WHERE { 
 ?statement rdf:subject dtsmm:microsoft . 
 ?statement rdf:predicate ?relation . 
 ?statement rdf:object ?object . 
}
\end{verbatim}
\end{footnotesize}
\end{tcolorbox}
\end{flushleft}
 \label{fig:Query4}
\end{figure}

which would produce exactly 48 statements representing new knowledge deriving from our KG. For example, looking at the acquire predicate type (i.e. \url{http://dtsmmkg.org/dtsmmkg/ontology#acquire}), we would know that Microsoft has acquired companies like \textit{Cloudknox\_Security}, \textit{CyberX} and \textit{RiskIQ}. In SPARQL we might then ask for information about this last: 


\begin{figure}[htp!]
\begin{flushleft}
\begin{tcolorbox}
\begin{footnotesize}
\begin{verbatim}
SELECT DISTINCT *
FROM <DTSMM_KG> 
WHERE { <http://dtsmmkg.org/dtsmmkg/resource/riskiq> ?p ?o . } 
\end{verbatim}
\end{footnotesize}
\end{tcolorbox}
\end{flushleft}
 \label{fig:Query5}
\end{figure}

with Turtle result as follows: 

\begin{figure}[htp!]
\begin{flushleft}
\begin{tcolorbox}
\begin{footnotesize}
\begin{verbatim}
@prefix dtsmm: <http://dtsmmkg.org/dtsmmkg/resource/> .
@prefix dtsmm-ont: <http://dtsmmkg.org/dtsmmkg/ontology#> .
@prefix owl: <http://www.w3.org/2002/07/owl#> .
@prefix skos: <http://www.w3.org/2004/02/skos/core#> .

dtsmm:cybersecurity_firm_riskiq a dtsmm-ont:Entity ;
 owl:sameAs <http://dbpedia.org/resource/RiskIQ> ;
 skos:related <http://dbpedia.org/resource/Computer_security> .
\end{verbatim}
\end{footnotesize}
\end{tcolorbox}
\end{flushleft}
 \label{fig:Query6}
\end{figure}


The results would tell us that this is a Computer Security company.
If we now would supply via RAG the existing 960 DBpedia knowledge triples on Microsoft plus the extracted relation triples deriving from our KG in-context to a LLM (in this example we used OpenAI GPT-4 Turbo\footnote{\url{https://platform.openai.com/docs/models/gpt-4-turbo-and-gpt-4}}), and then ask the question, specifying to be brief:\\

\textit{Is Microsoft dedicating substantial resources to computer security technologies?}\\

we would get the following answer from the system:\\

\textit{Yes, Microsoft is dedicating substantial resources to computer security technologies, as evidenced by its acquisitions of companies like RiskIQ, a leader in global threat intelligence and attack surface management, and CyberX, which specializes in securing IoT devices.}\\

where the latter information derives exactly from our KG, showing the power of the supplied new knowledge. 

In summary, when generating textual content, the RAG model can then reference the most recent updates and developments in Digital Transformation encapsulated within our knowledge graph. This ensures that the generated content is not only linguistically coherent but also factually accurate and up-to-date, reflecting the latest trends and information. Such enrichment is particularly valuable in applications where staying current with industry changes is critical, such as health-care, market analysis, or creating summaries for decision-makers. Our knowledge graph acts as a pool of novel knowledge taken from social media that RAG models can tap into, by supplying them with a repository of timely and relevant information.





\end{document}